# Coverage Performance in Aerial-Terrestrial HetNets

M. G. Khoshkholgh*, Keivan Navaie**, Halim Yanikomeroglu†, V. C. M. Leung*, and Kang. G. Shin††
* Department of Electrical and Computer Engineering, the University of British Columbia
** School of Computing and Communications, Lancaster University
† Department of System and Computer Engineering, Carleton University
†† Department of Electrical Engineering and Computer Science, The University of Michigan

*Abstract*—An efficient solution to improve the coverage in cellular networks is to use unmanned aerial vehicles (UAVs), augmented with the functionalities of terrestrial base stations (BSs). This paper investigates the coverage probability in multi-tier Aerial-Terrestrial HetNets, where in addition to the ground BSs (G-BSs), UAV-mounted BSs (U-BS) are also introduced across tiers to improve the coverage performance. We then model the Ground-to-Ground (G2G) and Air-to-Ground (A2G) links incorporating the impact of Line-of-Sight (LOS) and non-LOS (NLOS) path-loss attenuations in various wireless environments including sub-urban, urban, dense-urban, and high-rise. Using tools of stochastic geometry, we then obtain the coverage probability in such a setting and its upper-bound as a function of the percentage of U-BSs in each tier, as well as other system parameters. We use simulations to confirm the accuracy of our analysis and investigate the impact of various system parameters on the coverage probability. The thus-obtained upper-bound provides important quantitative insights on the network coverage design and the conflicting impacts of different system parameters. Our analysis also shows that in some communication environments, e.g., high-rise and dense urban, introducing U-BSs can degrade the coverage probability. Nevertheless, our analysis suggests that to minimize the coverage cost, one may consider turning off a given percentage of G-BSs. In urban and sub-urban areas, one can also adjust the altitude of U-BSs in order to improve the coverage probability. Such a strategy, however, is shown to be ineffective in dense-urban and high-rise environments.

## I. Introduction

Achieving universal connectivity via currently deployed terrestrial cellular networks is difficult due to severe path-loss attenuation, excessive inter-cell interference, and mismatch between hard-to-predict traffic demands and static infrastructure.

The use of Unmanned Areal Vehicles (UAV) has recently been proposed to enhance the coverage of Heterogenous cellular Networks (HetNets) [1, 2]. In such a network, a robust communication link is established between the base station installed in the UAV (also referred to as U-BSs) and the ground user equipments (UEs). In fact, the field measurements reported in [3, 4] confirm that compared to the Ground-to-Ground (G2G) link—the link between the ground BSs and the EUs—it is more likely for the link between the U-BS and the UE (Air-to-Ground (A2G)) to experience line-of-sight(LOS) propagation. Lower attenuation of LOS links enables more efficient connectivity while increasing the received interference for other users. Especially, use of U-BSs has been suggested to support UEs that experience severe shadowing and/or receive high interference. In such cases, the location of U-BSs becomes very important and a given coverage performance could be achieved by adjusting the density of U-BSs. The oOptimal altitude of the U-BSs for the maximum coverage has also been studied in [4]. An algorithm is also proposed in [5] for the optimal 3-D placement of U-BSs to maximize the coverage in cellular networks.

Using tools of stochastic geometry, [6] also provides practical insights on the coverage probability in the downlink of a single-tier UAV network without G-BSs. They further show that by raising the altitude of UAVs, both the coverage probability and spectral efficiency are decreased. Equally important is the communication between the G-BS and the UAVs. The coverage probability of the communication between the G-BS and UAVs is investigated in [7], incorporating the characteristics of an urban communication environment and considering the G-BSs' antenna tilt. The authors of [8] investigated a scenario in which a combination of drone UEs and ground UEs are served in a single-tier cellular networks. Their results show that the ground-to-air (G2A) communication link is prone to a high level of interference. The authors of [9] studied the coverage performance of on-demand UAV-assisted cellular networks and showed that the coverage performance is severely reduced for cases where the UAVs' altitude is higher than a threshold. In a UAV-enables LTE macro-cell system, the measurements reported in [10] also shows that for a U-BS with altitude of 150 m the link between U-BSs and ground UEs can suffer up to 7 dB SINR degradation compared to an equivalent ground communication link, due mainly to the interference. These results show the importance of interference which is directly related to the LOS/NLOS properties of the wireless link between/from the U-BSs.

In this paper, we investigate multi-tier aerial HetNet (A-HetNet) co-existing with traditional multi-tier terrestrial HetNets (T-HetNet). Multi-tier drone cellular networks are proposed in [2] as an effective way to guarantee the universal coverage and high transmission capacity. The authors of [11] investigate the benefits of multi-tier drone cellular networks over a single-tier network. The results in [11] are, however, based on the standard-path-loss model, which is not accurate for A2G and G2A links, see, [12, 13]. The coverage performance results in [2, 11] are also limited to simulations and lack a rigorous analytical framework.

Building upon analytical tools of stochastic geometry, and assuming Max-SIR cell association, we provide numerically tractable expressions for the coverage probability of the Aerial-Terrestrial HetNet (AT-HetNet). Our analysis shows that by increasing the density of U-BSs, the coverage probability declines; for the dense deployed networks the measured loss is substantially large. Our analysis also shows that the sub-urban environments suffer more severely than a high-rise setting, since the former is more susceptible to excessive LOS interference created by the U-BSs. One can moderately enhance the coverage probability of AT-HetNets by properly adjusting the altitude of U-BSs in sub-urban and urban environments.

Our system model is presented in Secion II, followed by coverage performance analysis in Section III. Section IV provides simulation and numerical results, and conclusions are drawn in Section V.

## II. SYSTEM MODEL

We focus on the downlink communication in an Aerial-Terrestrial Heterogeneous Cellular Network (AT-HetNet), comprising several classes/tiers of G-BSs and U-BSs. In this model, network consists of $K$ tiers of BSs (e.g., macro-cells, pico-cells, femto-cells, etc.), and each tier $i$, $i = 1, \ldots, K$, is characterized by $(\lambda_i, P_i, \beta_i, q_i)$, where $\lambda_i$ indicates BS's spatial density, $P_i$ is the BSs' transmission power, $\beta_i$ is the prescribed SIR threshold, and $q_i \in [0, 1]$ is the percentage of BSs in tier $i$ that are aerial (U-BS). Note that for $q_i = 0$ all the BSs in tier $i$ are G-BSs, and the system is reduced to the traditional HetNet system (T-HetNet). For $q_i = 1$, the BSs in the system are all aerial, namely Areal HetNet (A-HetNet), i.e., the terrestrial infrastructure does not exist or fails, for example, due to natural disasters. The system designer can then adjusts $q_i$ to achieve a given performance metric, e.g., coverage probability.

BSs at tier $i$ are spatially distributed via a homogeneous Poisson point process (PPP) $\Phi_i$ with given spatial density $\lambda_i \geq 0$. Assuming that each BS randomly chooses its status $s \in \{U, G\}$ where $U$ and $G$ stand for aerial and ground, respectively, we have $\Phi_i = \Phi_i^G \bigcup \Phi_i^U$, where $\Phi_i^G$ represents G-BSs of tier $i$ that is a PPP with density $(1 - q_i)\lambda_i$; $\Phi_i^U$ comprises U-BSs that is a PPP with density $q_i\lambda_i$; and $\Phi_i^G$ and $\Phi_i^U$ are independent for all $i$. UEs are single-antenna and distributed through a homogeneous PPP $\Phi_M$, independent of sets $\Phi_i$, with given spatial density $\lambda_M$. Using Slyvniak's theorem and network stationarity, one can evaluate the network performance for a given UE located at the origin, known as a *typical UE*.

In each tier, for simplicity we assume that the U-BSs also utilize the designated terrestrial spectrum band in the corresponding tier. Although there might be better spectrum allocation strategies, reusing the same spectrum band does not require major reconfiguration/upgrade in the corresponding tiers, thus facilitating deployment of the U-BSs.

Similar to [3, 10, 14] where off-the-shelf LTE/LTE-A radio equipments are used for quick and efficient deployment of the U-BSs, we assume that U-BSs are the exact replicas of G-BSs. Such an assumption also represents the trend in the wireless industry, see, e.g., [3, 14], and standardization activities in the 3rd Generation Partnership Project (3GPP) [15]. Using the above model, we investigate whether or not *the network coverage benefits from allowing the BSs to "fly."*

### A. Channel Model

Suppose U-BSs of tier $i$ are located at the same height, $H_i$, above the ground. To distinguish between G-BSs and U-BSs, we also assume that $H_i$ is usually larger than the antenna height in macro BSs, which is roughly $25 - -30$ m.

TABLE I
A2G PARAMETERS IN DIFFERENT WIRELESS ENVIRONMENTS [4].

|   | High-Rise | Dense-Urban | Urban | Sub-Urban |
|---|---|---|---|---|
| $\phi$ | 27.23 | 12.08 | 9.61 | 4.88 |
| $\psi$ | 0.08 | 0.11 | 0.16 | 0.43 |

The received signal at the typical UE originated from BS $x_i$ travels through LOS or NLOS channels, depending on its relative distance to the UE, density of buildings, environment, etc. For T-HetNets, several analytical models have been developed [12, 13, 16]. Regarding the A2G links, recent measurements also corroborate the existence of LOS/NLOS propagation modes [4, 10]. As a generic model, to include LOS/NLOS effects, we adopt the path-loss model recommended in the 3GPP [12, 13, 16], where for $s \in \{G, U\}$, the path-loss attenuation in tier $i$ is

$$L_i(\|x_i\|; s) = \begin{cases} L_L^i(\|x_i\|; s) & \sim p_L^i(\|x_i\|; s), \\ L_N^i(\|x_i\|; s) & \sim p_N^i(\|x_i\|; s). \end{cases} \quad (1)$$

For the G2G link, i.e., $s = G$ with $n_i \in \{L, N\}$, $L_{n_i}^i(\|x_i\|; G) = \phi_{n_i}^i(1 + \|x_i\|)^{-\alpha_{n_i}^i}$, where $\alpha_L^i$ (resp. $\alpha_N^i$) is the path-loss exponent associated with the LOS (resp. NLOS) link, $\phi_L^i$ (resp. $\phi_N^i$) is a constant, characterizing the LOS (NLOS) wireless propagation environment, and is related to various factors, e.g., the height of transceivers, antenna's beam width, weather, etc.

For the U-BSs, the path-loss function can be written as $L_{n_i}^i(\|x_i\|; U) = \psi_{n_i}^i(1 + \sqrt{\|x_i\|^2 + H_i^2})^{-\alpha_{n_i}^i}$, which depends on the height of BS $x_i$, $H_i$, and its $(x, y)$ distance to the origin, $\|x_i\|$.

In (1), for a BS located at position $x_i$ the probability of LOS mode is $p_{n_i}^i(\|x_i\|; s)$, where $\sum_{n_i \in \{L, N\}} p_{n_i}^i(\|x_i\|; s) = 1$. In this paper, for G2G communications we adopt the ITU-R UMi model [12, 16]

$$p_L^i(\|x_i\|; G) = \min\left\{\frac{D_0^i}{\|x_i\|}, 1\right\}\left(1 - e^{-\frac{\|x_i\|}{D_1^i}}\right) + e^{-\frac{\|x_i\|}{D_1^i}}, \quad (2)$$

where parameters $D_0^i$ and $D_1^i$ characterize the near-field (LOS) and far-field (NLOS) critical distances, respectively. Therefore, if $\|x_i\| \leq D_0^i$, BS $x_i$ is in the LOS mode.

Fig. 1-a shows the probability of LOS mode versus the distance for $\|x_i\| > D_0^i$. The probability of LOS mode is shown to decrease exponentially as the distance between the BS and the typical user increases.

The probability of a LOS channel between the U-BS $x_i \in \Phi_i^U$, and the typical receiver is [17]:

$$p_L^i(\|x_i\|; U) = \left(1 + \phi e^{-\psi\left(\frac{180}{\pi}\arctan(\frac{H_i}{\|x_i\|}) - \phi\right)}\right)^{-1}, \quad (3)$$

where $\phi$ and $\psi$ depend on the environment, e.g., height and density of buildings, urban or sub-urban areas, etc., see Table I. $\arctan(\frac{H_i}{\|x_i\|})$ is the elevation angel between the typical UE and the U-BS $x_i$. Increasing $H_i$ is shown to increase the probability of LOS mode, which is also shown in Fig. 1-a. Although increasing the altitude of the U-BSs improves the chance of LOS communication, it may increase the path-loss attenuation, thus reducing the received power at the typical user. Therefore, a higher LOS probability needs to be carefully balanced with an increase of path-loss attenuation [5].

In Fig. 1-a the LOS probabilities of G2G and A2G links are also compared. As expected A2G links provide a much higher LOS probability, particularly when the altitude of UAV gets higher. For a large enough distance between the transmitted and the typical user, we further note that the probability of LOS communication is close to 0, while for the A2G links (especially when $H$ is large enough), the LOS probability is much higher. Unlike the G2G links, the A2G links can, therefore, convey a much higher portion of the transmitted power to the receiver, almost independently of the distance.

### B. Inter-Cell Interference

The model we consider permits universal frequency reuse across all tiers. To model the interference, we assume that the typical UE is associated with BS $x$. In the next section, we further elaborate on the rules that associate UEs with BSs. The interference imposed by tier $j$, $I_j$, to the typical user is a shot noise:

$$I_j = \sum_{s \in \{U,G\}} I_j^s = \sum_{s \in \{U,G\}} \sum_{x_j \in \Phi_j^s \setminus x} P_j L_j(\|x_j\|; s) H_{x_j}^s \quad (4)$$

where $H_{x_j}^s$ is channel fading. In practice, the distribution of small-scale fading depends heavily on whether the link is LOS or NLOS. Note that the existence of fading in A2G links has been shown through measurements and simulations, e.g., in [10], due mainly to reflection, diffraction, and scattering caused by various objects in the environment. Here, we consider the following generic model to represent fading in our analysis:

$$H_{x_i}^s = \begin{cases} H_{x_i,L}^s \sim \Gamma(M_{i,L}^s, \frac{1}{M_{i,L}^s}) & \sim p_L(\|x_i\|; s) \\ H_{x_i,N}^s \sim \Gamma(M_{i,L}^s, \frac{1}{M_{i,L}^s}) & \sim p_N(\|x_i\|; s), \end{cases} \quad (5)$$

where $\Gamma(z, \frac{1}{z})$ is normalized gamma distribution with parameter $z$. For $z = 1$ and $z \to \infty$, $\Gamma(z, \frac{1}{z})$ becomes Rayleigh, and non-fading, respectively. We further assume that $M_{i,L}^s > M_{i,L}^s$. In fact, for an LOS channel, the received signal power tends to fluctuate less severely, and on average, the power loss will be smaller. In an NLOS channel, however, the received signal power fluctuates more severely, and on average loss is also higher.

Figs. 1-b and 1-c show the complementary cumulative distribution function (CCDF) of interference for several values of $q_2 = q_1 = q$. , Increasing $q$ is shown to significantly increase the the probability that the typical UE receives a much higher level of interference. This is consistent with the results in Fig. 1-a which shows a higher chance of being a LOS link in case of A2G link. This, compared to a G2G link, significantly increases the aggregated interference in A2G transmission. Therefore, although A2G links may result in higher power of the received signal, their impact might be canceled out as the receiver is likely to receive a very high aggregated interference thorough LOS links with the interferers.

### III. COVERAGE PROBABILITY

The typical UE successfully receives the signal transmitted by BS $x_i$, provided the corresponding SIR is larger than the prescribed signal-to-interference ratio (SIR) threshold, $\beta_i > 0$. The coverage probability is then equal to the CCDF of the SIR. Assuming that the typical UE is served by BS $x_i \in \Phi_i^s$, the SIR is

$$\text{SIR}_{x_i}^s = \frac{P_i L_i(\|x_i\|; s) H_{x_i}^s}{\sum_{j=1}^K I_j}. \quad (6)$$

In this paper, we consider Max-SIR cell association, where for each user, the BS (either U-BS or G-BS) providing the maximum SIR across all BSs is considered as the supporting BS. If there is at least one BS for which the typical UE can be successfully—with respect to the SIR threshold—associated with, i.e.,

$$\mathcal{A}^s = \left\{ \exists i : \max_{x_i \in \Phi_i^s, \forall i} \text{SIR}_{x_i}^s \geq \beta_i \right\} \neq \varnothing, \quad (7)$$

the typical UE is considered in the coverage area of the network. The probability of the typical UE, $p_{\text{cov},i}^s$, being associated with BS $x_i \in \Phi_i^s$ is

$$p_{\text{cov}} = \sum_{s \in \{U,G\}} \sum_{i=1}^K p_{\text{cov},i}^s. \quad (8)$$

In what follows, we derive an expression for $p_{\text{cov},i}^s$ as a function of system parameters.

Using the same line of argument as in our previous work [18], it is easy to show that the coverage probability, $p_{\text{cov},i}^s$, is upper-bounded as:

$$p_{\text{cov},i}^s \leq \mathbb{E} \sum_{x_i \in \Phi_i^s} \mathbf{1}\left(\text{SIR}_{x_i}^s \geq \beta_i\right)$$

$$= 2\pi \lambda_i^s \int_0^\infty x_i \mathbb{P}\left\{\text{SIR}_{x_i}^s \geq \beta_i\right\} dx_i, \quad (9)$$

where in the second step we use the Campbell-Mecke Theorem [19]. We also note that $\lambda_i^s = q_i \lambda_i \mathbf{1}_{s=U} + (1-q_i)\lambda_i \mathbf{1}_{s=G}$. We then write:

$$\mathbb{P}\left\{\text{SIR}_{x_i}^s \geq \beta_i\right\} = \mathbb{P}\left\{H_{x_i}^s \geq \frac{\beta_i \sum_{j=1}^K I_j}{P_i L_i^{n_i}(\|x_i\|; s)}\right\}$$

$$= \sum_{n_i \in \{L,N\}} p_i^{n_i}(\|x_i\|; s) \mathbb{P}\left\{H_{x_i,n_i}^s \geq \frac{\beta_i \sum_{j=1}^K I_j}{P_i L_i^{n_i}(\|x_i\|; s)}\right\}. \quad (10)$$

To evaluate the inner probability in (10), we adopt the Alzer's Lemma [20]:

$$\mathbb{P}\left\{H_{x_i,n_i}^s \geq \frac{\beta_i \sum_{j=1}^K I_j}{P_i L_i^{n_i}(\|x_i\|; s)}\right\}$$

$$\leq \mathbb{E}\left[1 - \left(1 - e^{-\nu_{i,n_i}^s \frac{M_{i,n_i}^s \beta_i \sum_{j=1}^K I_j}{P_i L_i^{n_i}(\|x_i\|; s)}}\right)^{M_{i,n_i}^s}\right]$$

$$= \sum_{l_{i,n_i}^s=1}^{M_{i,n_i}^s} \binom{M_{i,n_i}^s}{l_{i,n_i}^s} (-1)^{l_{i,n_i}^s+1} \mathcal{L}_I\left(\frac{\nu_{i,n_i}^s l_{i,n_i}^s M_{i,n_i}^s \beta_i}{P_i L_i^{n_i}(\|x_i\|; s)}\right), \quad (11)$$

where $\nu_{i,n_i}^s = M_{i,n_i}^s (M_{i,n_i}^s!)^{-1/M_{i,n_i}^s}$. Using (4), the Laplace transform of interference is

$$\mathcal{L}_I(t) = \mathbb{E}e^{-t \sum_{s \in \{G,U\}} \sum_{j=1}^K I_j^s} = \prod_{j=1}^K \prod_{s \in \{G,U\}} \mathbb{E}_{I_j^s}[e^{-tI_j^s}], \quad (12)$$

where

$$\mathbb{E}_{I_j^s}[e^{-tI_j^s}] = \left(\mathbb{E}_{\Phi_j^s} \prod_{x_j \in \Phi_j^s} \mathbb{E}e^{-tP_j L_j(s; \|x_j\|) h_{x_j}^s}\right)$$

$$= \mathbb{E}_{\Phi_j^s} \prod_{x_j \in \Phi_j^s} \left(\sum_{n_j \in \{L,N\}} \frac{p_{n_j}^j(s; \|x_j\|)}{(1 + tP_j L_{n_j}^j(s; \|x_j\|))^{M_{j,n_j}^s}}\right)$$

$$= e^{-2\pi \lambda_j^s \sum_{n_j} \int_0^\infty y_j p_{n_j}^j(s; y_j) \left(1 - \frac{1}{(1 + tP_j L_{n_j}^j(s;y_j))^{M_{j,n_j}^s}}\right) dy_j}. \quad (13)$$

Substituting (13) in (12), and the result into (11), we get:

$$\mathbb{P}\{\text{SIR}_{x_i}^s \geq \beta_i\} = \sum_{n_i \in \{L,N\}} \sum_{l_{i,n_i}^s=1}^{M_{i,n_i}^s} p_i^{n_i}(\|x_i\|; s) \frac{\binom{M_{i,n_i}^s}{l_{i,n_i}^s}}{(-1)^{l_{i,n_i}^s+1}}$$

$$\exp\left\{-2\pi \sum_{j=1}^K \sum_{s \in \{G,U\}} \lambda_j^s \sum_{n_j \in \{L,N\}} \int_0^\infty y_j p_{n_j}^j(s; y_j)\left(1 - \right.\right.$$

$$\left.\left.(1 + \frac{M_{i,n_i}^s \beta_i \nu_{i,n_i}^s l_{i,n_i}^s}{P_i L_i^{n_i}(\|x_i\|; s)} P_j L_j^{n_j}(y_j; s))^{-M_{j,n_j}^s}\right) dy_j\right\}. \quad (14)$$

Finally, substituting (14) into (9)

$$p_{\text{cov},i}^s = 2\pi \lambda_i^s \sum_{n_i \in \{L,N\}} \sum_{l_{i,n_i}^s=1}^{M_{i,n_i}^s} \binom{M_{i,n_i}^s}{l_{i,n_i}^s} (-1)^{l_{i,n_i}^s+1} \int_0^\infty x_i$$

$$p_i^{n_i}(x_i; s) \exp\left\{-2\pi \sum_{j=1}^K \sum_{s \in \{G,U\}} \lambda_j^s \sum_{n_j \in \{L,N\}} \int_0^\infty y_j p_{n_j}^j(s; y_j)\right.$$

$$\left.\left(1 - (1 + \frac{M_{i,n_i}^s \beta_i \nu_{i,n_i}^s l_{i,n_i}^s}{P_i L_i^{n_i}(x_i; s)} P_j L_j^{n_j}(y_j; s))^{-M_{j,n_j}^s}\right) dy_j\right\} dx_i.$$

The coverage probability can then be evaluated using (8).

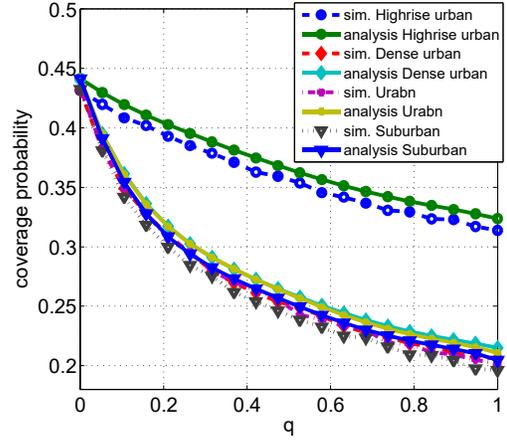

Fig. 2. Coverage probability vs. $q$, where $\lambda = 5 \times 10^{-3}$, $H = 200$m.

## IV. NUMERICAL AND SIMULATION RESULTS

For ease of exposition, we consider a single-tier scenario, $K = 1$, and drop tier index. The BSs are randomly distributed within a disk with radius $10,000$ units according to the corresponding tier densities (number of BSs per $km^2$). The BS transmit power, $P_2 = 1$W, and the LOS (resp. NLOS) path-loss exponent is $\alpha_1^L = \alpha_2^L = 2.4$ ($\alpha_1^N = \alpha_2^N = 4$), the path-loss intercept parameters are set to 1, $D^0 = 80$m, $D^1 = 164$m, and $\beta = 5$. We adopt the Monte Carlo technique and the presented results are based on analyzing $40,000$ simulation snapshots.

Fig. 2 shows the accuracy of our analysis and plots the obtained upper-bound and the simulation results vs. $q$ in different wireless environments. The thus-obtained upper-bound is shown to follow closely the simulations in all four simulated environments. Increasing $q$, i.e., increasing the percentage of U-BSs, is also shown to decrease the coverage probability. Using UAV-mounted BSs is shown to not necessarily improve the coverage probability.

For the case where all BSs are U-BS, $q = 1$, the coverage probability is decreased by $55\%$, where the decrease in the coverage probability is lowest (highest) in a high-rise (sub-urban) environment compared to other environments mainly because, in the high-rise environments, the existence of high density of blockages leads to dominant NLOS interfering links. In the case of sub-urban environment, many adjacent interfering U-BSs pose as LOS interfering links, substantially increasing interference.

Fig. reffig:CoveLambda shows the impact of density on the coverage probability for various values of $q$ in several prominent communication environments. In all cases, introducing U-BSs decreases the coverage probability. Especially in dense networks, $\lambda > 0.1$, using only U-BSs, i.e., $q = 1$, significantly reduces the network coverage performance regardless of the communication environment. This highlights the severe effect of aggressive interference in the UAV communication, and calls for sophisticated interference management in AT-

HetNets.

Fig. 3-a also suggests that $\lambda$ be carefully chosen to minimize the reduction in the coverage probability. For instance, in Fig. 3-a, by choosing $\lambda \approx 10^{-3}$, the coverage probability is slightly reduced for any $q > 0$. For density $\lambda_0 = 0.1$, one can turn off $1\%$ of all BSs randomly to reach density $\lambda_1 = 10^{-3}$. Furthermore, depending on feasibility, it is required to turn off $q\%$ of G-BSs and include $q\lambda_1$ U-BSs, without significantly reducing the coverage probability. Such a straightforward procedure may be appealing as it may reduce the energy cost of cellular networks, particularly when it is difficult to predict the traffic.

The impact of U-BSs' altitude on the coverage probability is investigated in Fig. 4. Figs. 4-a and 4-b suggest that for sub-urban and urban environments, one can increase the coverage probability by adjusting the U-BSs' altitude. The optimal value of altitude yielding the highest coverage probability is shown to be affected significantly by $q$. In fact, one should reduce the altitude by increasing $q$ in order to preserve the coverage gain. Figs. 4-a and 4-b further show that the gain made by optimally choosing the altitude of UAVs is higher in sub-urban environments. Nevertheless, in dense-urban environments, adjusting the altitude does not affect the coverage probability.

## V. Conclusions

An efficient solution to improve the coverage in cellular networks is to use unmanned aerial vehicles (UAVs), augmented with the functionalities of terrestrial base stations (BSs) in an Aerial-Terrestrial setting. In this paper, we have investigated the coverage probability in multi-tier Aerial-Terrestrial HetNets, where besides the ground BSs (G-BSs), UAV-mounted BSs (U-BS) are introduced across tiers to improve the coverage performance. We then modeled the Ground-to-Ground (G2G) and Air-to-Ground (A2G) links incorporating the impact of Line-of-Sight (LOS) and non-LOS (NLOS) path-loss attenuations in various wireless environments including sub-urban, urban, dense-urban, and high-rise. Adopting tools of stochastic geometry, we then derived the coverage probability in such a setting and its upper-bound as a function of the percentage of U-BSs in each tier, as well as other system parameters. Using simulations, we have confirmed the accuracy of our analysis and investigated the impact of various system parameters on the coverage probability. The thus-obtained upper-bound provides important quantitative insights on the network coverage design and the conflicting impacts of different system parameters. Our analysis has also shown that in some communication environments, such as high-rise and dense urban, introducing U-BSs can be detrimental to the coverage probability. Nevertheless, our analysis suggests that one may consider turning off a given percentage of G-BSs to minimize the coverage cost. Interestingly, for urban and sub-urban areas, one can further adjust the altitude of U-BSs in order to increase the coverage probability. Such a strategy, however, was shown ineffective in dense-urban and high-rise environments.

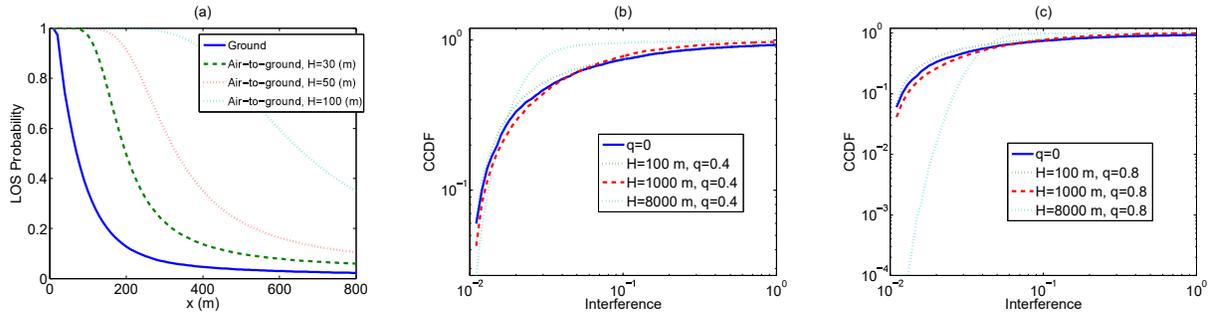

Fig. 1. (a) Comparison between LOS probability in G2G and A2G links. We only depict the LOS probability in tier 2, and set $D_0^2 = 18$ m and $D_1^2 = 36$ m.

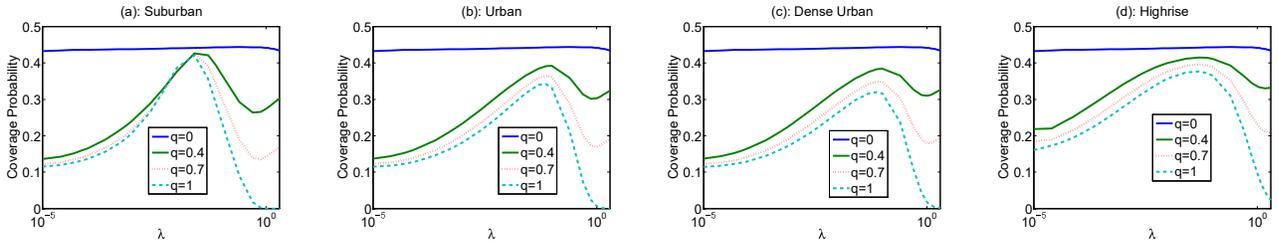

Fig. 3. Coverage probability versus $\lambda$ when $H = 400$ m.

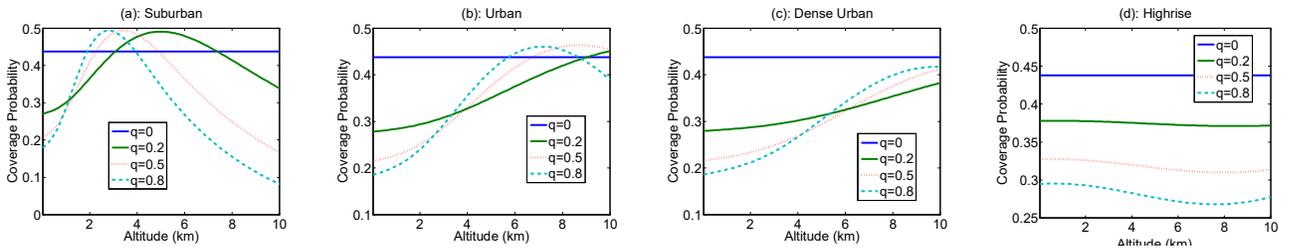

Fig. 4. Coverage probability versus $H$ when $\lambda = 5 \times 10^{-3}$.